\documentclass[11pt]{article}
\usepackage[margin=1in]{geometry}
\usepackage{graphicx}
\usepackage{dcolumn}
\usepackage{mathtools}
\usepackage{bm}
\usepackage{hyperref}
\usepackage{xcolor}
\usepackage[caption=false]{subfig}
\usepackage{subcaption}
\usepackage[numbers, sort&compress]{natbib}
\usepackage{cleveref}
\usepackage{rotating}
\usepackage{adjustbox}
\usepackage{tikz}
\usetikzlibrary{shapes}
\usepackage{amsmath}
\usepackage{amssymb}
\usepackage{booktabs}

\usepackage{etoolbox}
\usetikzlibrary{plotmarks}
\usepackage{MnSymbol}

\usepackage{comment}

\DeclareMathOperator{\Mellin}{\mathcal{W}}

\DeclareMathOperator{\dt}{\mathrm{d}t}
\DeclareMathOperator{\dk}{\mathrm{d}k}
\definecolor{darkgreen}{RGB}{0.0, 100, 0.0}
\definecolor{purple}{RGB}{128, 0.0, 128}
\definecolor{orange}{RGB}{255, 149, 0.0}
\definecolor{grey}{RGB}{127, 127, 127}

\definecolor{blueplot}{RGB}{12.0, 93, 165}
\definecolor{greenplot}{RGB}{0.0, 185, 69}
\definecolor{orangeplot}{RGB}{255, 149, 0.0}
\definecolor{brownplot}{RGB}{140, 86, 75}
\definecolor{pinkplot}{RGB}{227, 119, 194}

\usetikzlibrary{shapes.geometric}

\usepackage{lipsum}

\title{\bfseries Stability of inverse cascade in NLS Wave turbulence}

\author{G. Costa\textsuperscript{1\dag}, G. Krstulovic\textsuperscript{1}, S. Nazarenko\textsuperscript{1}}

\date{}

\begin{document}

\maketitle
\vspace{-1.5em}  
\begin{center}
    \textsuperscript{1}Université Côte d’Azur, CNRS, Institut de Physique de Nice (INPHYNI), Nice.\\
    \vspace{0.5em}
    \textsuperscript{\dag} \textit{Corresponding author:} \texttt{guillaume.costa@bbox.fr}
\end{center}
\maketitle
\vspace{-1em}

\begin{abstract}
    We study the stability of the inverse cascade in the 3D  Nonlinear Schrödinger Wave Kinetic Equation (WKE) in the presence and absence of a condensate. Using theoretical arguments supported by numerical simulation, we show that, the Kolmogorov-Zakharov solution is stable with respect to small isotropic perturbations in the absence of a condensate. Conversely, introducing a condensate component leads to a growth of the condensate occupation and ultimately destabilizes the inverse cascade. The system then evolves towards a new steady-state whose properties depend on the condensate dissipation rate.
\end{abstract}

\section{Introduction}
    The Nonlinear Schrödinger (NLS) equation governs the dynamics of a wide variety of physical systems ranging from nonlinear optics to plasma physics and Bose-Einstein condensates. In the weakly nonlinear regime, statistical properties of wave fields governed by the NLS equation can be studied within the framework of weak wave turbulence theory (WWT), through the wave kinetic equation (WKE)~\cite{dyachenko1992optical, deng2023full}. 
    
    The WKE governs the dynamics of the wave amplitude spectrum, called waveaction, predicting the existence of thermal equilibrium, Rayleigh-Jeans (RJ) and nonequilibrium Kolmogorov-Zakharov (KZ) stationary solutions associated with cascades of conserved quantities. For the three-dimensional NLS equation, theoretical predictions indicate the existence of both an inverse cascade of wave action and a direct cascade of energy~\cite{nazarenko2011wave, dyachenko1992optical}. The WKE exhibits finite-time blowup, a phenomenon that is not observed in the dynamics of the 3D NLS equation. This discrepancy stems from the fact that the WKE does not capture the formation of a condensate, which plays a crucial role in the NLS dynamics. To quantitatively account for this effect and extend the class of admissible solutions, one can allow for a non-zero mass concentrated at $\bm{k}=\bm{0}$~\cite{semikoz1995kinetics}.

    While stationary solutions of the NLS-WKE have been widely studied~\cite{proment2009quantum, griffin2022energy, falkovich1988energy, pushkarev2000turbulence, zhu2023direct, zhu2023self, labarre2025wave, brazhnikov2001capillary, abdurakhimov2007turbulence, falcon2009capillary}, the stability properties of these solutions, especially in the presence of coherent structures such as condensates, remain less understood. Indeed, the stability of these solutions has been the subject of a few rigorous studies~\cite{escobedo2007fundamental, collot2024stability}, only in the absence of condensates. When the inverse cascade transports sufficient wave action toward the lowest wavenumbers, the system may develop a macroscopic occupation of the ground mode, leading to the formation of a condensate. The presence of such a condensate significantly modifies the wave interaction dynamics and may lead to qualitatively different equilibria.

    Despite substantial progress in the theory of wave turbulence for the NLS equation, the stability of KZ solutions under perturbations and their interaction with a condensate remain important open questions. In particular, it is unclear how small perturbations evolve in the presence or absence of a condensate and how this affects the long-time dynamics of the inverse cascade.

    In this work, we investigate the stability of the inverse-cascade KZ solution of the 3D NLS equation in the presence and in the absence of a condensate. We combine theoretical predictions based on a linear stability analysis of the WKE with numerical simulations of the forced-dissipated NLS equation. We show that, in the absence of a condensate, the system converges toward the KZ solution predicted by the WWT and that this solution is linearly stable with respect to small isotropic perturbations. In contrast, the presence of a condensate renders the KZ solution linearly unstable: perturbations grow exponentially, ultimately driving the system toward the RJ equilibrium.
    
\section{Theoretical background around the WKE}
    The NLS equation
    \begin{equation*}
        \begin{split}
            &i \Psi_t + \Delta \Psi \pm |\Psi|^2\Psi = 0, \\
            &\mathcal{H} = \int \left(|\bm{\nabla}\Psi|^2 \mp |\Psi|^4 / 2 \right)\mathrm{d}\bm{x},
        \end{split}
    \end{equation*}
    describes the evolution of the wavefunction $\Psi(\bm{x},t)$ in a conservative system of weakly nonlinear dispersive waves. It arises in many physical contexts including optical fibers~\cite{agrawal2012nonlinear}, Bose-Einstein condensates~\cite{dalfovo1999theory}, plasma physics~\cite{zakharov1972collapse}, and the envelope dynamics of deep-water gravity waves~\cite{hasimoto1972nonlinear}.
    When the quadratic part of the Hamiltonian $\mathcal{H}$ dominates the quartic one, the NLS equation admits a wave turbulence description, leading to the establishment of a kinetic equation~\cite{dyachenko1992optical, deng2023full}, for the wave-action spectrum $n_k = \left(\frac{2\pi}{L}\right)^3\langle |\Psi(\bm{x},t)|^2 \rangle$
    \begin{equation}
        \begin{split}
            \partial_t n_k = S_4[n_k] =  \frac{32\pi^3}{k} \int &\sqrt{\min\left(\omega_k, \omega_1, \omega_2, \omega_3 \right)}\; n_1n_2n_3n_k \left(\frac{1}{n_k} + \frac{1}{n_1} - \frac{1}{n_2} - \frac{1}{n_3}\right)\delta \omega^{k,1}_{2,3}\mathrm{d}\omega_1 \mathrm{d}\omega_2\mathrm{d}\omega_3,
        \end{split}
        \label{eq:WKE}
    \end{equation}
    where $\delta\omega^{k,1}_{2,3} = \delta\left(\omega_k  + \omega_1 - \omega_2 - \omega_3\right)$, and $\omega_k$ is given by the dispersion relation $\omega_k = k^2$. Note that in this work, collision integrals will be written in $\omega$-variable as it allows for an easier integration.
    
    The WKE~\eqref{eq:WKE} possesses two quadratic invariants, the number of particles and the energy
    \begin{equation}
        N = 4\pi \int k^2n_k \dk,  \qquad  H =  4\pi \int k^2 \omega_k n_k \dk, 
    \end{equation}
    whose conservation gives rise to an inverse wave-action cascade and a direct energy cascade. Stationary solutions of the WKE can be found by performing the Zakharov transform~\cite{zhu2023direct}, yielding two types of steady-state solutions. The system admits an equilibrium solution, the RJ thermal equilibrium spectrum $n_k = \dfrac{T}{\omega_k + \mu}$, corresponding to a zero-flux solution. In contrast, when the system is maintained out of equilibrium through forcing and dissipation, the system develops the KZ spectra 
    \begin{equation}
        n_k = C_{\mathrm{KZ}}^Q |Q_0|^{1/3}k^{-7/3}, \quad n_k =  C_{\mathrm{KZ}}^PP_0^{1/3}k^{-3}\ln^{-1/3}\left(k/k_f \right),
    \end{equation}
    associated with the inverse waveaction and direct energy cascade~\cite{zhu2023direct}. Here, $Q_0$ and $P_0$ denote the waveaction and energy fluxes, respectively, $k_f$ is the forcing scale, and $C_{\mathrm{KZ}}^Q = 7.577 \times 10^{-2}$ $C_{\mathrm{KZ}}^P = 5.26 \times 10^{-2}$ are the Kolmogorov-Zakharov constants.

    To account for the condensate formation and to extend the class of admissible solutions, we decompose the wave action into a regular component and a singular contribution at zero wavenumber
    \begin{equation}
        n_{\bm{k}}(t) = \tilde{n}_{\bm{k}}(t) + \delta(\bm{k})n_c(t),
    \end{equation}
    where $\tilde{n}_k$ represents the bulk spectrum and $n_c$ the condensate amplitude~\cite{semikoz1995kinetics}.
    
    Substituting into the WKE~\eqref{eq:WKE} (see Appendix~\ref{sec:Condensate}) yields the coupled equations
    \begin{align}
        \partial_tn_c &= -4\pi n_c \int k^2 S_3\left[\tilde{n}_k \right]\dk  = -n_c Q_3\left[\tilde{n}_k\right], \\
        \partial_t \tilde{n}_k &= S_4[\tilde{n}_k] +  n_c S_3[\tilde{n}_k], 
    \end{align}
    where $S_4$ is the four-wave collision integral defined in Eq.~\eqref{eq:WKE}, and $S_3$ a three-wave collision integral given by
    \begin{equation}
        \begin{split}
            S_3\left[\tilde{n}_k\right] &= \frac{8\pi^2}{k}\int k_2k_3\left[\mathcal{N}^{k}_{2,3} - 2 \mathcal{N}^{3}_{k,2}\right]\dk_2\dk_3, \\
            \mathcal{N}^{k}_{2,3} &= \delta\omega^k_{2,3} \left[\tilde{n}_2\tilde{n}_3 - \tilde{n}_k\left(\tilde{n}_2 + \tilde{n}_3 \right)\right].
        \end{split}
    \end{equation}
    While the condensate does not contribute to the energy budget, it enters the particle-number budget. Consequently, the unforced, non dissipative system possesses the following quadratic invariants
    \begin{equation}
        N_{\text{tot}} = \tilde{N} + n_c, \qquad H =  4\pi \int k^2 \omega_k \tilde{n}_k \dk, 
    \end{equation}
    where $\tilde{N} =  4\pi \int k^2 \tilde{n}_k \dk$ is the number of particles in the bulk. The addition of a condensate gives rise to an extra three-wave interaction term, so that the resulting equation contains both the original four-wave and a new three-wave contribution. While the four-wave equation admits KZ stationary solutions associated with constant fluxes, the three-wave term breaks the homogeneity of the equation, leaving the RJ distribution as the only exact steady state.

    In this work, we investigate the stability of the inverse wave action cascade, both in the absence and in the presence of a condensate. Without a condensate, the inverse cascade is expected to be stable, as it has been observed both numerically and experimentally~\cite{zhu2023direct,moreno2025observation}. Our aim is to characterize the relaxation towards this non-equilibrium steady state. We then investigate how the presence of a condensate, through the additional three-wave interactions, modifies this picture and drives the dynamics towards the RJ equilibrium.
        
\section{Numerical methods}
    \begin{figure}[!htb]
        \centering
        \includegraphics[width=1.0\textwidth]{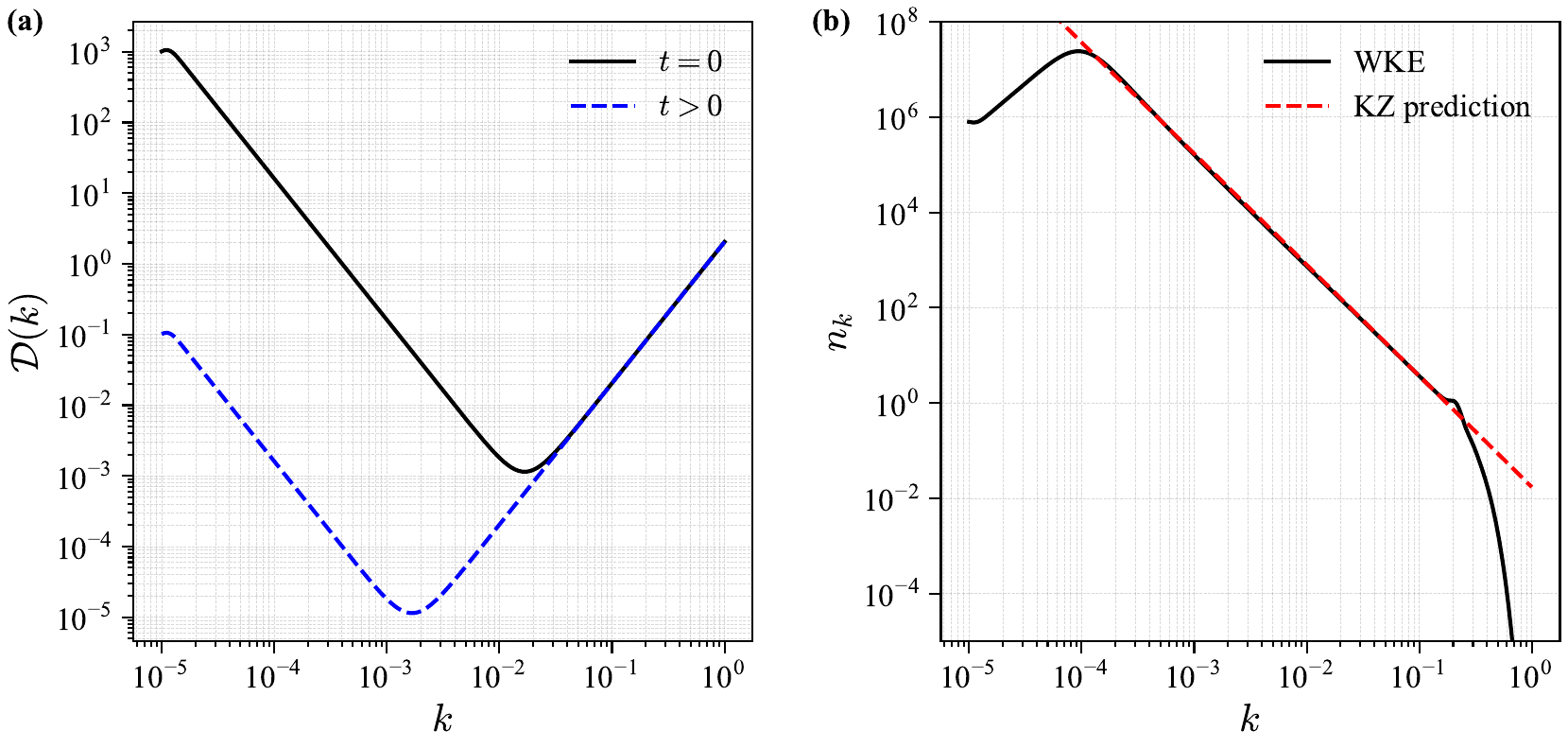}
        \caption{\textbf{(a)} Snapshot of the damping form to obtain the largest possible inertial range for the KZ spectrum (see footnote). The black dashed line corresponds to the initial damping with $k_d = k_d(0)$, while the dashed blue line corresponds to a damping at time $t>0$, where $k_d(t) < k_d(0)$. \textbf{(b)} Waveaction spectrum for the inverse waveaction cascade (Run 1, Tab.~\ref{tab:numdet}). The red dashed line represents the KZ solution $n_k= C_{\mathrm{KZ}}^Q |Q_0|^{1/3}k^{-7/3}$, with $C_{\mathrm{KZ}}^Q = 7.577 \times10^{-2}$~\cite{zhu2023direct}.}
        \label{fig:KZ}
    \end{figure}
    Using WavKinS~\cite{krstulovic2025wavkins}, we solve the WKE~\eqref{eq:WKE} on a logarithmic grid, with wave numbers following a geometric progression $k_n = k_0\lambda^n$. The grid parameter $\lambda$ is imposed by the resolution so that $\lambda(N, k_0, k_{\max}) = (k_{\max} / k_0)^{1/N}$. As mentioned previously, the WKE admits two classes of steady-state solutions depending on the presence of external excitations. To model such effects, we introduce both a forcing term $f(k)$ and a damping term $\mathcal{D}(k)$ to the bulk, so that the WKE becomes
    \begin{align}
        \partial_tn_c &= -n_c \left(\int k^2 S_3\left[\tilde{n}_k \right]\dk + \mathcal{D}_c\right)  = -n_c \left(Q_3\left[\tilde{n}_k\right]+\mathcal{D}_c\right), \label{eq:eq_cond} \\
        \partial_t \tilde{n}_k &= S_4[\tilde{n}_k] +  n_c S_3[\tilde{n}_k] + f(k) - \mathcal{D}(k), \label{eq:bulk}
    \end{align}
    where $\mathcal{D}_c$ is the condensate damping, and the forcing and damping operators are defined as 
    \begin{align}
        \mathcal{D}(k) &= \left[\left(\dfrac{k}{k_\nu}\right)^2 +\left(\dfrac{k_d}{k}\right)^2\varphi(k)\right] n_k, \\
        f(k) &= k^4 \exp\left[-\frac{\left(k-k_f\right)^2}{2(\Delta k_f)^2}\right],
    \end{align}
    with ($k_\nu$, $k_d$) denoting the small and large dissipation scales, and ($k_f$, $\Delta k_f$) the forcing position and width. The function
    \begin{equation}
        \varphi(k) = \left(1-\exp\left[-\left(k/\Delta k_d\right)^5\right]\right)
    \end{equation}
    is set to impose a finite dissipation in the limit $k \to 0$\footnote{In practice, to prepare the KZ solution with the largest inertial range allowed by our simulation, we first set the IR dissipation wavenumber to a relatively large value and wait for the system to reach a steady state. This procedure prevents the finite-time blow-up of the 3D NLS-WKE~\cite{semikoz1995kinetics,zhu2023direct}. We then gradually shift the IR dissipation wavenumber towards smaller values until the target value is reached.}.
    \begin{table}[!htb]
        \centering
        \begin{tabular}{|c||c|c|c|c|c|c|c|c|}
            \hline
             & $N$ & $k_{\min}$ & $k_{\max}$ & $k_f$ & $\Delta k_f$ & $k_{\nu}$ & $k_d$ & $\Delta k_d$ \\
            \hline\hline
            Run 1 & $1600$ & $10^{-7}$ & $1.0$ & $0.2$ & $2\times 10^{-2}$ & $0.7$ & $7 \times 10^{-6}$ & $+\infty$\\
            Run 2  & $800$ & $10^{-5}$ & $1.0$ &  $0.2$ & $2\times 10^{-2}$ & $0.7$ & $4 \times 10^{-4}$ & $10^{-6}$\\
            \hline
        \end{tabular}
        \caption{Numerical details of the base simulations, reaching the KZ steady-state. These simulations are then used as initial conditions for perturbed runs. Run 1 is used without condensate while Run 2 allows for finite dissipation in the $k \to 0$ limit, consistent with the existence of a condensate.}
        \label{tab:numdet}
    \end{table}
    To study the stability of the inverse cascade, we first perform numerical simulations of the KZ steady-state (Fig.~\ref{fig:KZ}b), whose parameters are reported in Tab.~\ref{tab:numdet}. Note that the parameters and forcing are set to maximize the inertial range for the inverse cascade. These simulations are then perturbed either by adding a small isotropic perturbation (Run 1) or a small condensate (Run 2).

\section{Stability of solutions in the absence of condensate}
    While the stability of steady-states solutions of the WKE has only been tackled in a few studies, a comprehensive theory of stability has been developed by Balk and Zakharov~\cite{balk1998stability}, that relies on the use of the Mellin transform. The resulting equation involves a special function $\Mellin$ that contains all the information regarding the stability of the system. For further information, the authors refer the reader to~\cite{balk1998stability} for the theoretical development, and to~\cite{costa2026stability} for an application to the stability of acoustic waves.
    
    We first study the linear stability of the KZ solution $n_k \sim k^{-7/3}$, associated with the inverse cascade, in the absence of condensate (i.e. $n_c = 0)$. 
    \subsection{Linearized equation - Carleman equation}
        Let $n_k^0$ be a steady-state solution of the WKE~\eqref{eq:WKE} and $A_k \ll 1$ a small isotropic perturbation such that the waveaction reads
        \begin{equation}
            n_k = n_k^0(1 + A_k), \qquad n_k^0 = Bk^{-7/3} = C\omega^{-7/6}.
        \end{equation}
        Linearizing Eq.~\eqref{eq:WKE} around $n_k^0$, and keeping only leading order term yields
        \begin{equation}
            \begin{split}
                \partial_tA_k &= \frac{32\pi^3}{kn_k^0}\int \sqrt{\min(\omega_k, \omega_1,\omega_2,\omega_3)} \left(\mathcal{A}^{1,k}_{2,3}+ \mathcal{A}^{k,1}_{2,3} - 2\mathcal{A}^{2,3}_{1,k}\right)\delta\omega^{k,1}_{2,3}\mathrm{d}\omega_1\mathrm{d}\omega_2\mathrm{d}\omega_3, \\
                \mathcal{A}^{k,1}_{2,3} &= A_kn_k^0\left[n_2^0n_3^0 - n_1^0 \left(n_3^0 +  n_2^0\right)\right].
            \end{split}
        \end{equation}
        This equation can be simplified by taking the Mellin transform, formally a Laplace transform in log-coordinates, defined for any complex-valued function $f$ as $ \mathcal{M}[f](s) = \int_0^{+\infty} \omega^{s-1} f(\omega) \, d\omega, \; \forall s \in \mathbb{C}$, yielding the Carleman equation:
        \begin{equation}
            \begin{split}
                \partial_t \hat{A}(s + h) &= \Mellin(s) \hat{A}(s) + \Psi(s+h), \\
                \hat{A}(s)& = \mathcal{M}[A_k](s), 
            \end{split}
        \end{equation}
        where $h = 1/3$ is the homogeneity of the linearized collision integral, $\Psi$ contains all the details of the forcing and initial conditions of the perturbation, and $\Mellin$ is the Mellin function, defined as follows
        \begin{equation}
            \begin{split}
                &\Mellin(s) = 4\pi^3\int \sqrt{\min(1, q_1,q_2,q_3)} \left[P_{s}(q_1, q_2, q_3) - 2 Q_{s}(q_1,q_2,q_3) \right]\delta(1+q_1-q_2-q_3)\mathrm{d}q_1\mathrm{d}q_2\mathrm{d}q_3,\\
                &P_{s}(q_1,q_2,q_3) = q_1^{-s-7/6} \left(q_2^{-7/6}q_3^{-7/6} - q_2^{-7/6} - q_3^{-7/6} \right) + q_2^{-7/6}q_3^{-7/6} - q_1^{-7/6}\left(q_2^{-7/6} + q_3^{-7/6} \right), \\
                &Q_{s}(q_1,q_2,q_3) = q_2^{-s-7/6} \left(q_1^{-7/6} - q_3^{-7/6} - q_1^{-7/6}q_3^{-7/6} \right).
            \end{split}
            \label{eq:Mellin}
        \end{equation}
        As stated in~\cite{balk1998stability, zakharov2012kolmogorov, costa2026stability}, the stability analysis is reduced to the study of the properties of the Mellin function $\Mellin$ and, in particular, of its winding number $\kappa$ about the origin, defined as 
        \begin{equation}
            \label{eq:Winding}
                    \kappa(\sigma) = \frac{1}{2\pi i}\oint_\mathcal{C} \frac{dz}{z},
                    \qquad\mathcal{C} = \big\{ \mathcal{W}(\sigma + i\delta) \;|\; \delta \in \mathbb{R} \big\},
        \end{equation}
        where the contour $\mathcal{C}$ is the image of the vertical line $\sigma + i\delta$ under the map $\mathcal{W}$, with $\delta \in \mathbb{R}$ (see Figure~\ref{fig:MellinProp}a).
        More precisely, the stability is directly linked to the existence of a zero-winding strip $\Pi\left(\sigma_-, \sigma_+ \right)$ and to the relative position of the peculiar point $s=0$ with respect to this strip. For $h>0$, the stability criterion reduces to
        \begin{equation}
            \label{eq:stabcrit}
            \sigma_-  \leq 0 \leq \sigma_+ + h.
        \end{equation}
            
    \subsection{Mellin function and predictions}
        \begin{figure}[!htb]
            \centering
            \includegraphics[width=1.\textwidth]{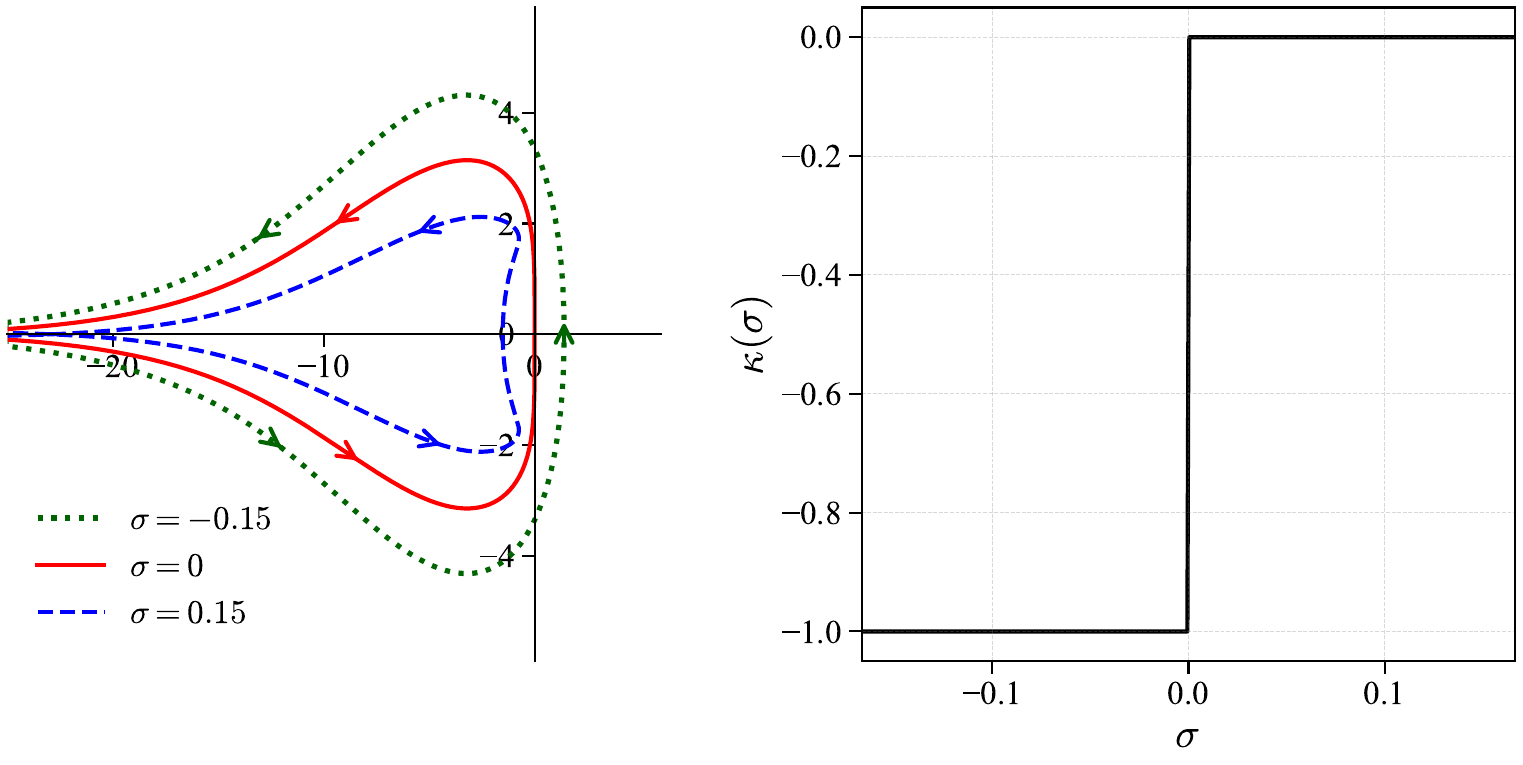}
            \put(-472, 240){\resizebox{14pt}{!}{\textbf{(a)}}}
            \put(-250, 240){\resizebox{14pt}{!}{\textbf{(b)}}}                
            \put(-310, 240){\resizebox{32pt}{!}{$\mathcal{I}m(\Mellin)$}}
            \put(-290, 120){\resizebox{32pt}{!}{$\mathrm{Re}(\Mellin)$}}
            \caption{Properties of the Mellin function. \textbf{(a)} Paths $\mathcal{C} = \big\{ \mathcal{W}(\sigma + i\delta) \;|\; \delta \in \mathbb{R} \big\}$ for different $\sigma = \mathrm{Re}(s)$. \textbf{(b)} Winding number $\kappa$ of the Mellin function $\Mellin$ as a function of $\sigma = \mathrm{Re}(s)$.}
            \label{fig:MellinProp}
        \end{figure}
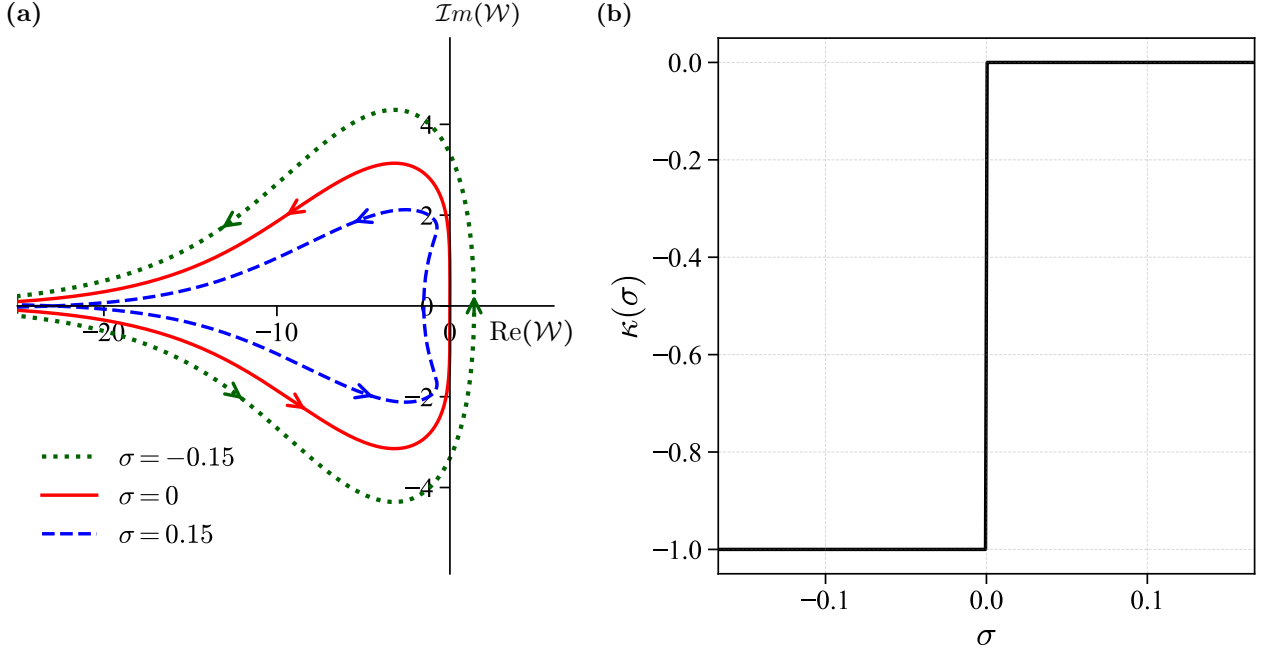
        The computation of the winding number $\kappa$~\eqref{eq:Winding} relies on the calculation of the Mellin function~\eqref{eq:Mellin}, which can be expressed using hypergeometric functions. The expression of the Mellin function $\Mellin$ is given in Appendix~\ref{sec:Mellin}\footnote{A similar expression can be found in~\cite{escobedo2007fundamental}, using the series development of the hypergeometric functions $_3F_2$.}. The Mellin function $\Mellin$, is analytic in the strip $\Pi(-4/3, 1/6)$ and has a unique zero within this strip, located at $s = 0$. Consequently, the sub-strip $\Pi(0, 1/6)$ is the zero-winding strip (see Fig.~\ref{fig:MellinProp}). Based on the criterion~\eqref{eq:stabcrit}, the KZ solution associated with the inverse cascade is stable w.r.t isotropic perturbations. The perturbation can be expressed as a propagating wave in the positive $x = \ln{\left(\omega_k/\omega_p\right)}$ direction (as $h>0$) with an exponential envelope $e^{-x/6}$, corresponding to the slowest decay, and possesses the following self-similarity
        \begin{equation}
            A(x, t\tau) = A(x+3\ln{\tau}, t)e^{-x/6}, \qquad \left(x = \ln{\left(\omega_k/\omega_p\right)}\right).
        \end{equation}
        The perturbation can thus be rewritten as:
        \begin{align}
            &\ 
            \left\{\begin{aligned}
                     &A(x,t) = A(x_\tau, \tau)e^{-x/6}, \\
                     &x = x_\tau +3\ln{(t/\tau)},
                    \end{aligned}\right.
            \text \; \forall t \geq \tau > 0.
            \label{eq:Selfsim}
        \end{align}
        \begin{figure}[!htb]
            \centering
            \includegraphics[width=1\columnwidth]{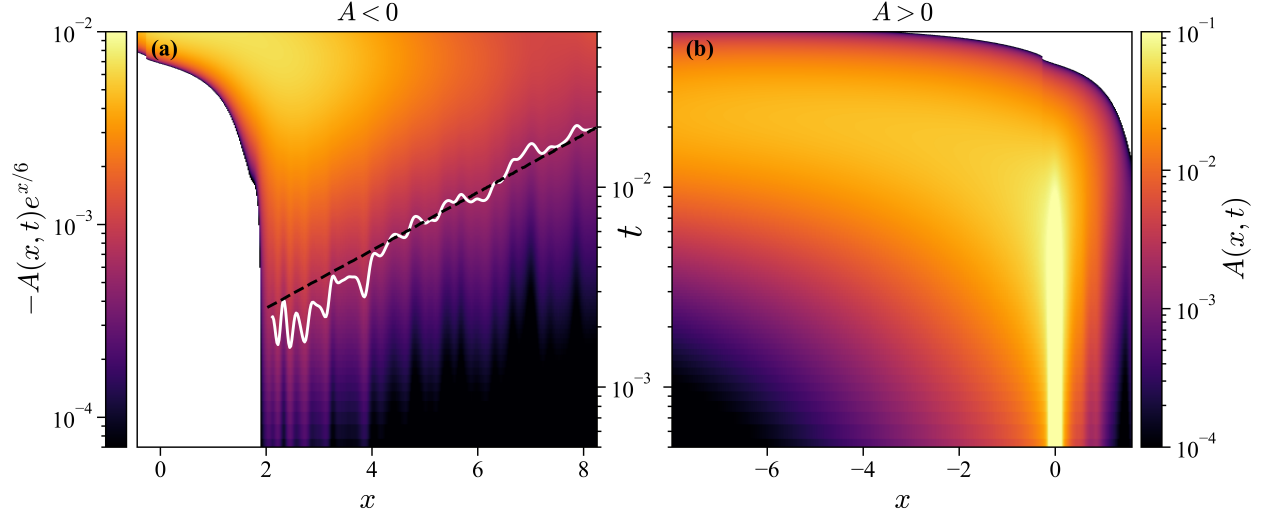}
            \caption{\textbf{(a)} $x-t$ diagram of the compensated perturbation $A(x,t)e^{x/6}$ highlighting a propagation of the perturbation towards small scales (large $x$) in agreement with the theoretical prediction of Eq.~\eqref{eq:Selfsim}. The solid white line is associated with a constant amplitude $A(x,t)e^{x/6} = const.$ compatible with the predicted path $x(t) = x_\tau + 3\ln\left(t/\tau\right)$ (black dashed line). \textbf{(b)} $x-t$ diagram of the perturbation with positive amplitude, carried by the inverse cascade.}
            \label{fig:Heatmap}
        \end{figure}
        Fig.~\ref{fig:Heatmap}a presents the $x-t$ diagram of the compensated perturbation $A(x,t)e^{x/6}$ highlighting the propagation of the perturbation, with negative amplitude, towards the positive $x$-direction, in agreement with the predictions of Eq.~\eqref{eq:Selfsim}. Note that the numerical simulations are performed on the full equation, implying the existence of nonlinear effects as the perturbation grows, ultimately leading to a backward propagation, with positive amplitude, associated with the inverse cascade (Fig.~\ref{fig:Heatmap}b). The authors would like to add a remark about the necessary condition for stability in the isotropic case mentioned in~\cite{zakharov2012kolmogorov}, namely $\Mellin'(0) < 0$. This condition is satisfied in our case (see Appendix~\ref{sec:Mellin}, Fig.~\ref{fig:Mellin_real}), and corresponds to a solution with a flux in the correct direction (i.e. an inverse cascade). However, this condition is  necessary but not sufficient, which means that the stability result is truly nontrivial.

        In the absence of a condensate, the KZ solution is thus stable with respect to small isotropic perturbations. However, the 3D NLS dynamics allows for the formation of a condensate whose effects must be taken into account.
        
\section{Stability of solutions in the presence of condensate}
    We now investigate the stability of the KZ solutions in the presence of a condensate. The dynamics of the system are described by Eqs.~\eqref{eq:eq_cond}~\&~\eqref{eq:bulk}.
    The dynamics of the condensate is obtained solving equation~\eqref{eq:eq_cond}, which admits the formal solution
    \begin{equation}
        n_c(t)=n_c(0)\exp\left[-\int_0^tQ_3[\tilde n_k]\,d\tau-\mathcal D_ct\right].
    \end{equation}
    The evolution of the condensate is therefore entirely determined by the sign of the three-wave flux $Q_3$. Neglecting dissipation, $Q_3<0$ leads to an exponential growth of the condensate, whereas $Q_3>0$ causes its decay.
    
    To determine the sign of $Q_3$, we consider a power-law bulk spectrum $\tilde n_k= B k^{-x}$, for which the three-wave collision integral reads
    \begin{equation}
        \begin{split}
            S_3[\tilde n_k] &= 8\pi^2B^2 k^{1-2x}I(x),\\
            I(x) &= \frac{2}{x-2}
            -\frac{\sqrt{\pi}\,2^{x-3}
            \left(\sec\left(\frac{\pi x}{2}\right)-1\right)
            \Gamma\left(1-\frac{x}{2}\right)}
            {\Gamma\left(\frac{3-x}{2}\right)}.
        \end{split}
        \label{eq:scaling}
    \end{equation}
    The function $I(x)$ was computed in~\cite{zhu2026turbulence} in the context of short Bogoliubov waves, for which the collision integral is exactly given by $S_3$.
    \begin{figure}[!htb]
        \centering
        \includegraphics[width=.5\textwidth]{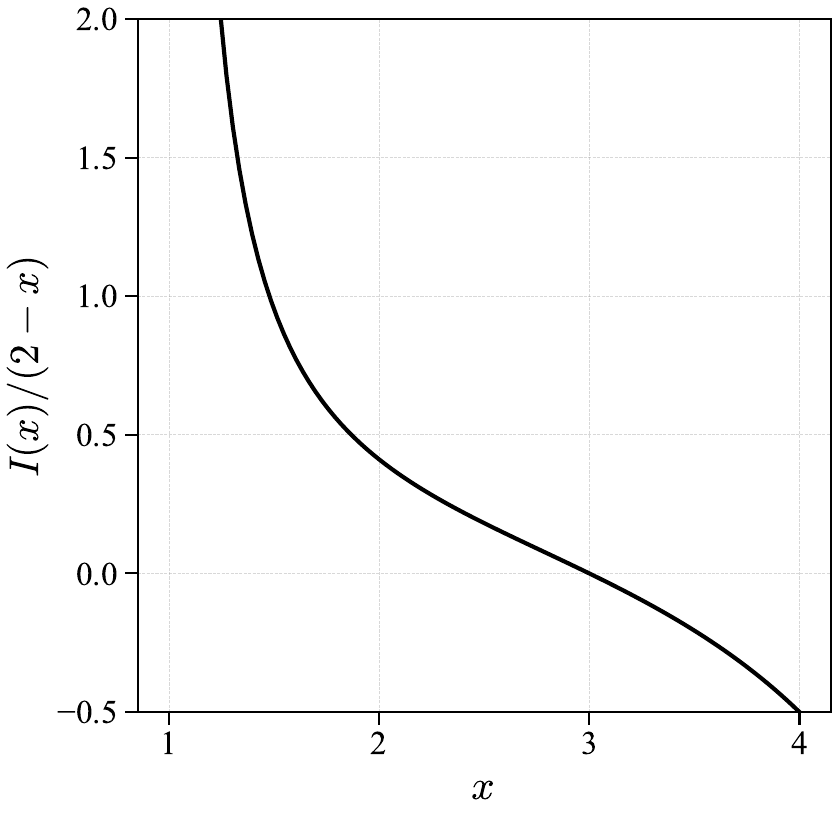}
        \caption{Scaling factor $I(x) / (2-x)$ of the three-wave flux $Q_3$, highlighting a change of sign for $x = 3$.}
        \label{fig:Isign}
    \end{figure}
    The corresponding three-wave flux is obtained by integrating the collision integral over wavevector space
    \begin{equation}
        Q_3[\tilde n_k] =\int S_3[\tilde n_k]\,d\mathbf{k} =4\pi\int_{k_{\min}}^{k_{\max}}k^2S_3[\tilde n_k]\,\dk,
    \end{equation}
    where $k_{\min}$ and $k_{\max}$ denote the IR and UV cutoffs, roughly corresponding to the dissipation and forcing wavenumbers, respectively. Substituting Eq.~\eqref{eq:scaling} into the above expression yields
    \begin{equation}
        Q_3[\tilde{n}_k] = 16\pi^3\frac{I(x)}{2-x}k_{\min}^{3-2x} \left[\left(\frac{k_{\min}}{k_{\max}}\right)^{2x-3}-1\right].
    \end{equation}
    Since the term in square brackets is negative for $x > 3/2$ and positive for $x < 3/2$, combining its sign with that of $I(x)/(2-x)$, shown in Fig.~\ref{fig:Isign}, yields $Q_3[\tilde{n}_k]<0$ for $3/2 <x <3$ and $Q_3[\tilde{n}_k] \geq 0$ otherwise.
    
    In particular, the KZ inverse-cascade solution is characterized by the exponent $x=7/3$, which belongs to the interval $(3/2,3)$. Consequently, $Q_3[\tilde n_k]<0$, implying an exponential growth of the condensate through Eq.~\eqref{eq:eq_cond}. This result suggests that the KZ inverse-cascade solution becomes unstable in the presence of a condensate.

    \subsection{Undamped condensate dynamics}
        We first consider the case without condensate dissipation, $\mathcal D_c=0$. As previously shown, the KZ inverse-cascade spectrum satisfies $Q_3<0$, implying that the condensate is expected to grow.
        
        \begin{figure}[!htb]
            \centering
            \includegraphics[width=1.\textwidth]{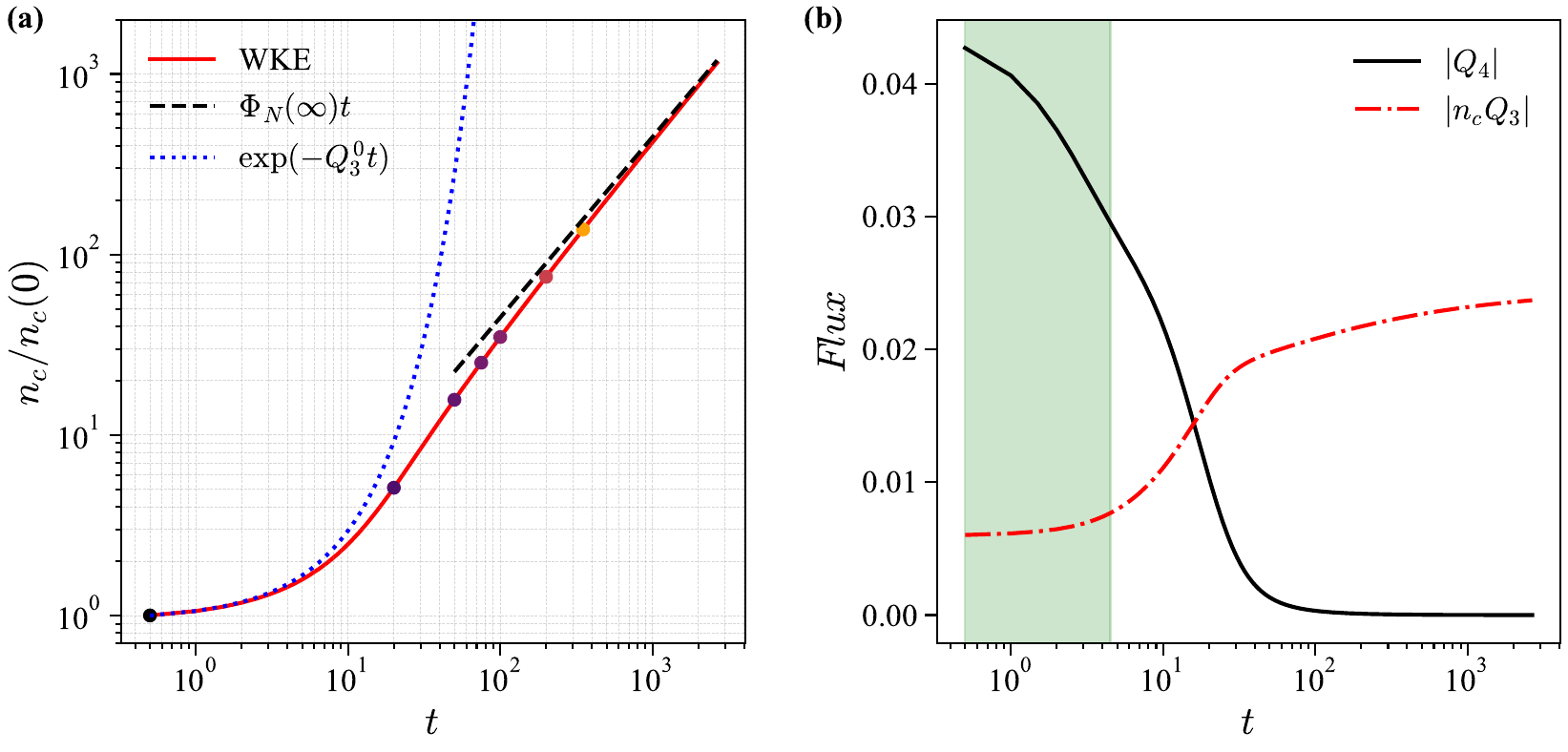}
            \caption{\textbf{(a)} Time evolution of the condensate highlighting an exponential increase at small times (dotted line) followed by a linear evolution compatible with the predictions of Eq.~\eqref{eq:nclongt} (dashed line). \textbf{(b)} Comparison of the three- and four-wave fluxes, indicating the dominant regime. The green shaded area represents the regime $n_cS_3[\tilde n_k]\ll S_4[\tilde n_k]$.}
            \label{fig:Regime}
        \end{figure}
        At early times, the condensate remains sufficiently weak that its contribution to the bulk dynamics is negligible,
        \begin{equation*}
            n_cS_3[\tilde n_k]\ll S_4[\tilde n_k].
        \end{equation*}
        The bulk spectrum therefore stays close to the KZ solution, so that the three-wave flux $Q_3$ is approximately constant, equal to $Q_3^0$. Equation~\eqref{eq:eq_cond} then predicts an exponential growth of the condensate,
        \begin{equation}
            n_c(t) \approx n_c(0) \exp\left[-Q_3^0 t\right],
        \end{equation}
        which is confirmed numerically in Fig.~\ref{fig:Regime}a.
        
\begin{figure}[!htb]
            \centering
            \includegraphics[width=1.\textwidth]{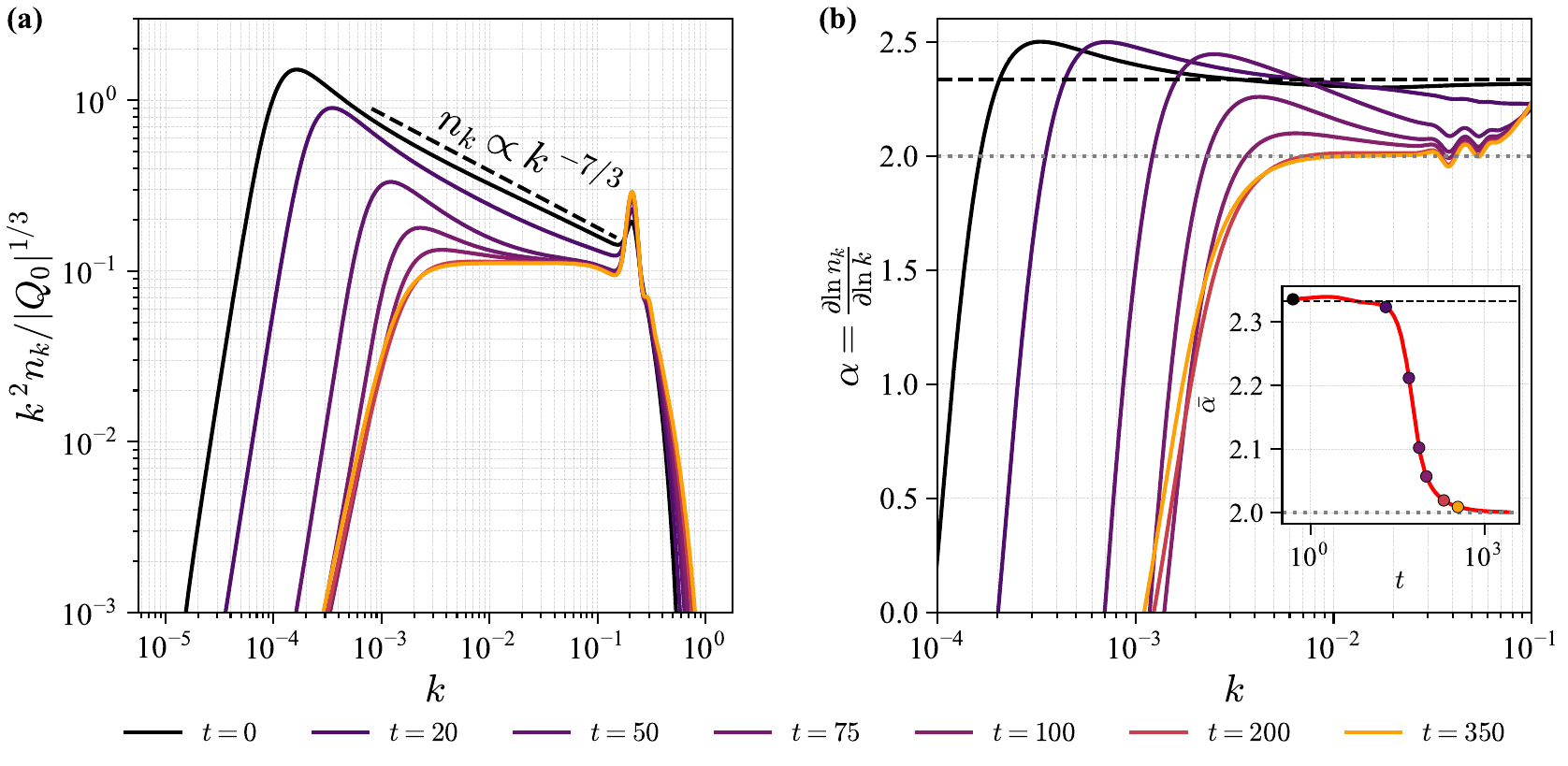}
            \caption{\textbf{(a)} Time evolution of the compensated wave-action spectrum, highlighting a transition from the KZ scaling, $n_k\propto k^{-7/3}$, RJ scaling, $n_k\propto k^{-2}$, induced by condensate formation. \textbf{(b)} Evolution of the local spectral exponent $\alpha$, showing the gradual crossover from the KZ value, $\alpha=7/3$, to the RJ value, $\alpha=2$. The inset displays the time evolution of the average spectral exponent $\bar{\alpha}$ over the inertial range.}
            \label{fig:Condensat}
        \end{figure}

        This regime persists until the nonlinear contributions become comparable, i.e., when $n_c S_3 \sim S_4$, corresponding to $t \approx 4.5$ in our simulations (Fig.~\ref{fig:Regime}b, green shaded area). Beyond this point, the condensate begins to significantly affect the bulk spectrum. This interaction manifests as a thermalization front propagating from small to large scales, as illustrated in Fig.~\ref{fig:Condensat}a.
        As time progresses, the spectral slope evolves continuously: the local power-law exponent in $k^{-\alpha}$ decreases from the KZ value $\alpha = 7/3$ toward the RJ equilibrium value $\alpha = 2$ reaching a new thermal equilibrium steady-state (Fig.~\ref{fig:Condensat}b).
        
        In the long-time limit, the bulk spectrum approaches thermal equilibrium, for which the four-wave particle flux vanishes. Integrating Eq.~\eqref{eq:bulk} over wave vectors gives the evolution of the total number of particles contained in the bulk,
        \begin{equation}
            \frac{\mathrm{d}\tilde N}{\dt}=n_cQ_3+4\pi\int k^2(f-\mathcal D)\,\dk=-\frac{\mathrm{d}n_c}{\dt}+\Phi_N\left[\tilde{n}\right],
        \end{equation}
        where $\Phi_N$ denotes the external flux of particles. 
        \begin{figure}[!htb]
            \centering
            \includegraphics[width=0.5\linewidth]{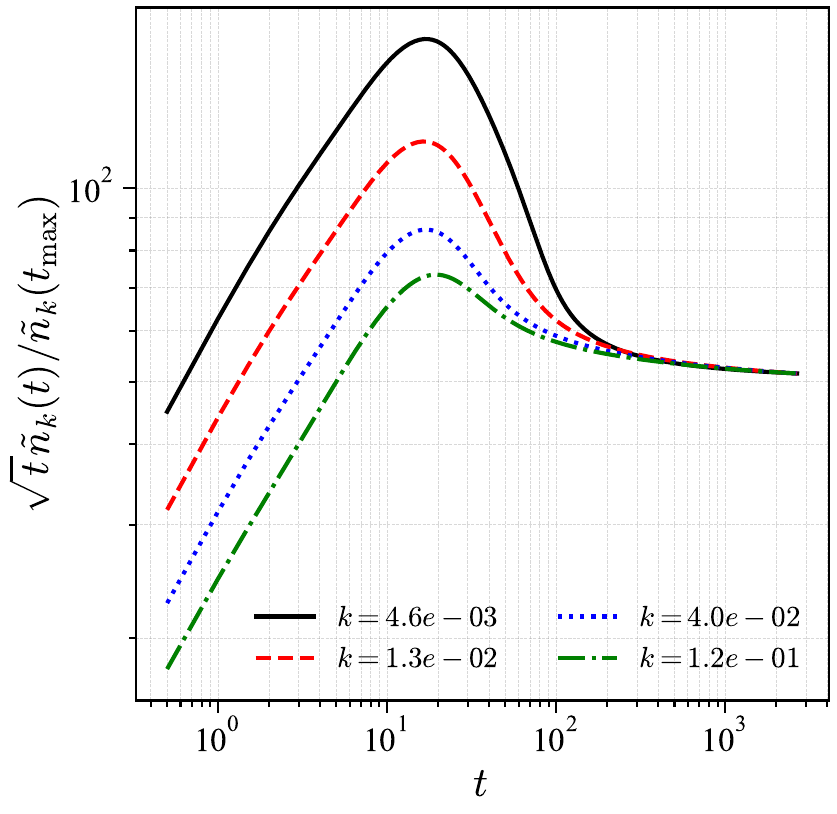}
            \caption{Compensated time evolution of the bulk spectrum $\tilde{n}_k$ at various wavenumber, confirming the prediction of Eq.~\ref{eq:bulk_t}.}
            \label{fig:bulk_vs_t}
        \end{figure}
        
        At late times, the bulk particle number $\tilde{N}$ varies slowly compared with the condensate amplitude
        \begin{equation*}
            \left|\frac{\mathrm{d}\tilde N}{\dt}\right| \ll \left|\frac{\mathrm{d}n_c}{\dt}\right|,
        \end{equation*}
        indicating that the injected particles are predominantly transferred to the condensate. Consequently, the condensate grows asymptotically at a constant rate,
        \begin{equation}
            n_c(t)\underset{t\to\infty}{\approx}\Phi_N t,
            \label{eq:nclongt}
        \end{equation}
        in excellent agreement with the numerical results shown in Fig.~\ref{fig:Regime}a (dashed line). Substituting this asymptotic behavior into Eq.~\ref{eq:eq_cond} (setting $\mathcal{D}_c = 0$) yields 
        \begin{equation}
            Q_3(t)  \sim t^{-1} \quad \Leftrightarrow \quad\tilde{n}_k(t) \sim t^{-1/2},
            \label{eq:bulk_t}
        \end{equation}
        which is confirmed by our numerical simulations. The corresponding compensated time evolution of $\tilde{n}_k$ at different wavenumbers is shown in Fig.~\ref{fig:bulk_vs_t}.
        
    \subsection{Effect of condensate dissipation}
     We now investigate the effect of a finite condensate dissipation, $\mathcal D_c>0$. The strength of the dissipation is characterized by the dimensionless ratio $\mathcal{D}_c/\epsilon$, where $\epsilon$ denotes the injection rate. 
        \begin{figure}[!htb]
            \centering
            \includegraphics[width=1.\textwidth]{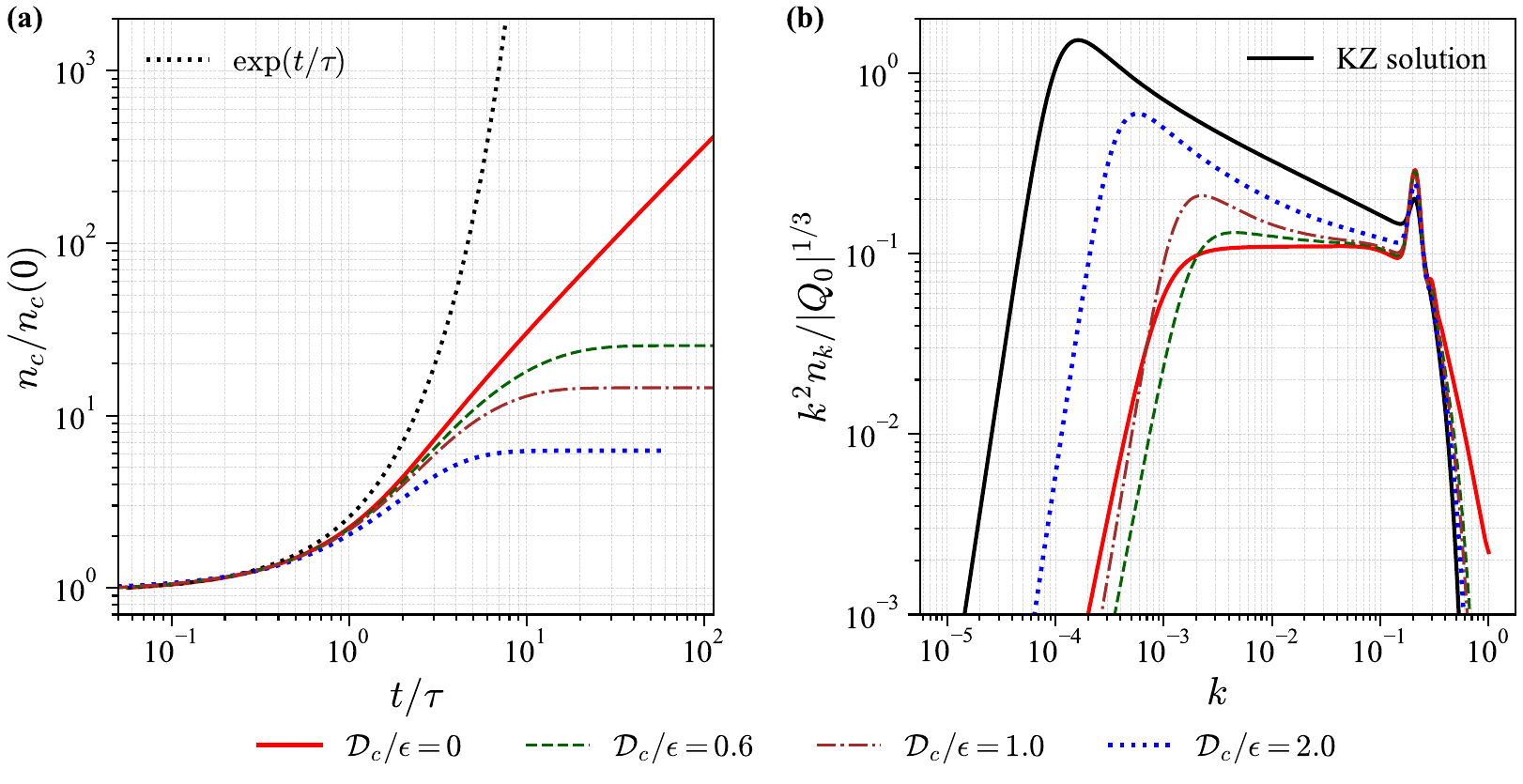}
            \caption{\textbf{(a)} Time evolution of the condensate for various dissipation values highlighting the upper bound $\mathcal{N}_c$. \textbf{(b)} Equilibrium spectrum for the various dissipation highlighting the coexistence of the two steady-state solutions: RJ (small scales) and KZ (large scales) plus a crossover regime.}
            \label{fig:Diss_Condensat}
        \end{figure}
        
        The condensate evolution is governed by the competition between the three-wave transfer from the bulk and the dissipative loss,
        \begin{equation}
            \partial_t n_c=-n_c(Q_3(t)+\mathcal D_c).
        \end{equation}
        
        For strong dissipation $\mathcal{D}_c > Q_3^0$, the condensate vanishes exponentially, leading to the standard KZ solution. For moderate dissipation, $\mathcal{D}_c \lesssim Q_3^0$, the early time dynamics, for which $Q_3(t) \gg \mathcal{D}_c$, follows the same exponential behavior as the undamped condensate case, albeit for a modified characteristic time $\tau$
        \begin{equation*}
            n_c(t) \approx n_c(0) \exp\left[t/\tau\right], \qquad \tau = -(Q_3^0 + \mathcal{D}_c)^{-1},
        \end{equation*}
        as illustrated by the early-time collapse of Fig.~\ref{fig:Diss_Condensat}a.

        \begin{figure}[!htb]
            \centering
            \includegraphics[width=1.\textwidth]{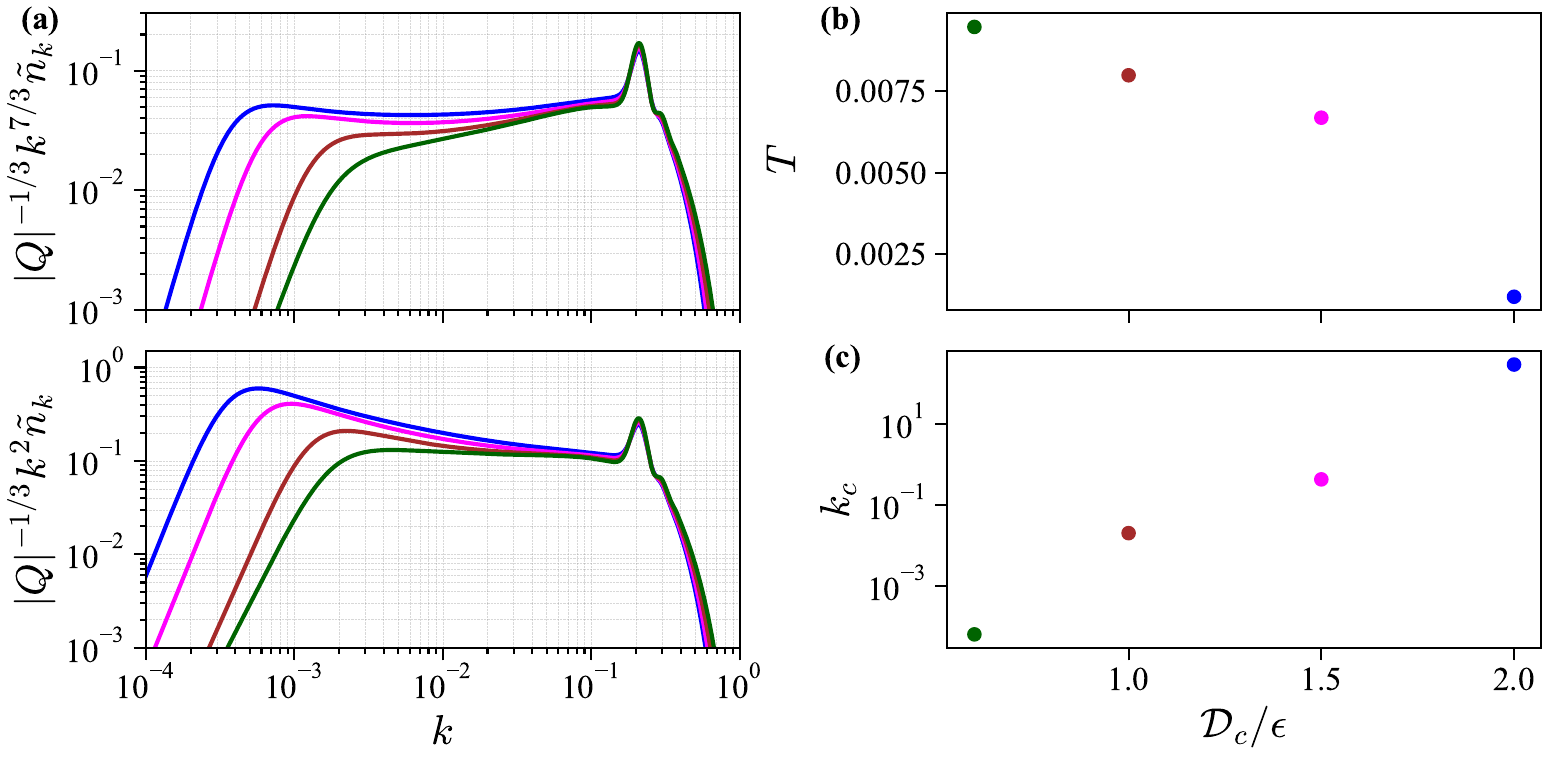}
            \caption{\textbf{(a)} New steady state for the bulk-spectrum $\tilde{n}_k$ compensated by the KZ (\textit{top panel}) and RJ (\textit{bottom panel}) slopes, highlighting the co-existence of the two regimes in agreement with the predictions of Eq.~\ref{eq:ansatznk}. \textbf{(b-c)} Extracted temperature $T$ and crossover wavenumber $k_c$ supporting the scenario of halted thermalization wave.}
            \label{fig:Steady-state}
        \end{figure}
        As time progresses, a new equilibrium is reached when $Q_3(t) = \mathcal{D}_c$, setting an upper bound $\mathcal{N}_c$ for the condensate amplitude (Fig.~\ref{fig:Diss_Condensat}a). As mentioned in the previous section, the thermalization occurs under the form of a front, propagating from large to small wavenumbers that is now halted by the presence of dissipation.
        To approximate this new equilibrium spectrum, we introduce the natural ansatz
        \begin{equation}
            \tilde{n}_k = Ak^{-7/3} + Bk^{-2},
            \label{eq:ansatznk}
        \end{equation}
        corresponding to a crossover between the KZ and RJ solutions, well observed in Fig.~\ref{fig:Diss_Condensat}b (dashed line). Assuming $\omega_k=k^2$, for which the RJ spectrum reads $n_k^{\mathrm{RJ}}\approx T k^{-2}$, we identify $B\equiv T$. Substituting into Eq.~\eqref{eq:ansatznk} and introducing the crossover wavenumber $k_c$ such that $Ak_c^{-7/3} = Bk_c^{-2}$ yields
        \begin{equation}
            \tilde{n}_k = T k^{-2}\left[1+ \left(\frac{k}{k_c}\right)^{-1/3}\right].
            \label{eq:New_state}
        \end{equation}
        Using least mean square regressions, we confirm the validity of the ansatz~\eqref{eq:New_state} (Fig.~\ref{fig:Steady-state}a) and extract the values of $(T,\,k_c)$ (Fig.~\ref{fig:Steady-state}b-c), highlighting the progressive emergence of the RJ thermal equilibrium as $\mathcal{D}_c/\epsilon$ decreases.

\section{Conclusion}
    In this work, we have investigated the linear stability of the 3D NLS-WKE inverse cascade in the presence and absence of a condensate. Using theoretical arguments, supported by numerical simulations, we have shown that the stability of the inverse cascade KZ solution is tied to the presence of a condensate. In the absence of a condensate, the inverse cascade solution is stable with respect to small perturbations. The relaxation to equilibrium occurs under the form of a wave-packet traveling in the opposite direction of the cascade, i.e. towards small scales, with an exponentially decreasing envelope $\exp\left[-x/6\right]$ matching the theoretical predictions. 
    
    The addition of a small, and mildly damped condensate ($D_c < Q_3^0$) initially leaves the inverse cascade solution unaffected. As time progresses, however, the condensate amplitudes grows, first exponentially, then either saturates at a plateau ($\mathcal{D}_c >0$), or continues to grow linearly ($\mathcal{D}_c = 0$). In this regime, the interactions between the bulk spectrum and the condensates becomes significant, leading to deviations from the KZ solution. These deviations take the form of a thermalization wave, propagating from small to large scales. Depending on the condensate dissipation strength $\mathcal{D}_c$, the system reaches a new steady-state characterized by a strong condensate and a partially thermalized bulk $\tilde{n}_k = T k^{-2}\left[1+ \left(\dfrac{k}{k_c}\right)^{-1/3}\right]$.
    
\section*{Acknowledgment}
     This work was funded by the Simons Foundation Collaboration grant Wave Turbulence (Award ID 651471). This work was supported by the French government through the France 2030 investment plan managed by the National Research Agency (ANR), as part of the Initiative of Excellence Université Côte d’Azur under reference number ANR-15-IDEX-01. The authors are grateful to the Université Côte d’Azur’s Center for High-Performance Computing (OPAL infrastructure) for providing resources and support.

\bibliographystyle{unsrtnat}  
\bibliography{apssamp}

\appendix
\section{The Mellin function}
    \label{sec:Mellin}
    The Mellin function for the 3D NLS inverse cascade is defined $\forall s\in \mathbb{C} \,/\,  \operatorname{Re}(s) \in\left]-4/3, 1/6 \right[$, as
    \begin{equation*}
            \begin{split}
                &\Mellin(s) = 4\pi^3\int \sqrt{\min(1, q_1,q_2,q_3)} \left[P_{s}(q_1, q_2, q_3) - 2 Q_{s}(q_1,q_2,q_3) \right]\delta(1+q_1-q_2-q_3)dq_1dq_2dq_3,\\
                &P_{s}(q_1,q_2,q_3) = q_1^{-s-7/6} \left(q_2^{-7/6}q_3^{-7/6} - q_2^{-7/6} - q_3^{-7/6} \right) + q_2^{-7/6}q_3^{-7/6} - q_1^{-7/6}\left(q_2^{-7/6} + q_3^{-7/6} \right), \\
                &Q_{s}(q_1,q_2,q_3) = q_2^{-s-7/6} \left(q_1^{-7/6} - q_3^{-7/6} - q_1^{-7/6}q_3^{-7/6} \right).
            \end{split}
    \end{equation*}
    \begin{figure}[!htb]
        \centering
        \resizebox{0.5\textwidth}{!}{%
        \begin{tikzpicture}
            \tikzstyle{every node}=[font=\normalsize\selectfont]
            \fill[yellow!20]
                (5.625,12.375) --
                (5.625,16.375) --
                (11.875,16.375) --
                (11.875, 10.375) --
                (7.625,10.375) --
                cycle;
            \draw [->] (5.625,10.375) -- (5.625,16.375);
            \draw [->] (5.625,10.375) -- (11.875,10.375);
            \draw [-] (5.625,12.375) -- (7.625,10.375);
            \draw [dashed] (5.625,12.375) -- (11.875,12.375);
            \draw [dashed] (7.625,10.375) -- (7.625,16.375);
            \node [font=\normalsize, inner xsep=0.080cm, inner ysep=0.085cm, rounded corners=0.020cm] at (6.9583,11.7083) {$\mathcal{D}_1$};
            \node [font=\normalsize, inner xsep=0.080cm, inner ysep=0.085cm, rounded corners=0.020cm] at (6.625,14.375) {$\mathcal{D}_2$};
            \node [font=\normalsize, inner xsep=0.080cm, inner ysep=0.085cm, rounded corners=0.020cm] at (9.75,14.375) {$\mathcal{D}_4$};
            \node [font=\normalsize, inner xsep=0.080cm, inner ysep=0.085cm, rounded corners=0.020cm] at (9.75,11.375) {$\mathcal{D}_3$};
            \node [font=\normalsize, inner xsep=0.080cm, inner ysep=0.085cm, rounded corners=0.020cm] at (12.375,10.375) {$q_2$};
            \node [font=\normalsize, inner xsep=0.080cm, inner ysep=0.085cm, rounded corners=0.020cm] at (5.25,16.5) {$q_3$};
            \node [font=\normalsize, inner xsep=0.080cm, inner ysep=0.085cm, rounded corners=0.020cm] at (7.625,10.1) {1};
            \node [font=\normalsize, inner xsep=0.080cm, inner ysep=0.085cm, rounded corners=0.020cm] at (5.35,12.375) {1};
        \end{tikzpicture}
        }%
        \caption{Decomposition of the integration domain for the Mellin function.}
        \label{fig:domains}
    \end{figure}
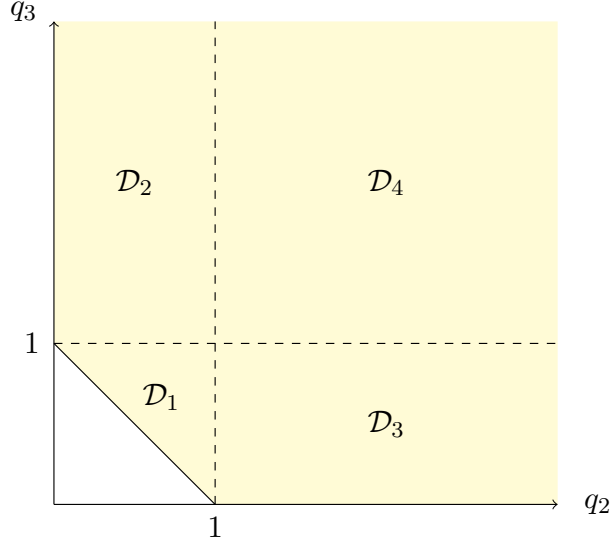
    
    The integral can be decomposed into four domains (Fig.~\ref{fig:domains}) which can be remapped onto the small triangle $\mathcal{D}_1$ by performing the Zakharov transform~\cite{nazarenko2011wave, zhu2023direct}, yielding
    \begin{equation*}
        \begin{aligned}
            \Mellin(s)&= 4\pi^3\int_{\mathcal{D}_1} \sqrt{q_1}\,\delta(1+q_1-q_2-q_3)\,\bigl(1+q_1^s-q_2^s-q_3^s\bigr) \\
            &\quad \times
            \Bigl[F_s(q_1,q_2,q_3,1)+F_s(1,q_2,q_3,q_1)-F_s(q_2,q_1,1,q_3)-F_s(q_3,q_1,1,q_2)\Bigr]\,\mathrm{d}q_1\,\mathrm{d}q_2\,\mathrm{d}q_3,
        \end{aligned}
    \end{equation*}
    where $F$ is given by
    \begin{equation*}
        F_s(x,y,z,t) = x^{-s-7/6}
        \Bigl[(yz)^{-7/6}- t^{-7/6}\bigl(y^{-7/6}+z^{-7/6}\bigr)\Bigr].
    \end{equation*}
    $\forall s\in \mathbb{C} \,/\,  \operatorname{Re}(s) \in\left]-4/3, 1/6 \right[$, the above integral is piece-wise convergent such that one can split the Mellin function into a sum of integrals of the form
    \begin{equation*}
        I(\alpha, \beta, \gamma) = \int_{\mathcal{D}_1}\delta(1+q_1-q_2-q_3) q_1^{\alpha}q_1^{\beta}q_3^{\gamma}\mathrm{d}q_1\mathrm{d}q_2\mathrm{d}q_3 = \frac{1}{(\alpha + \beta + 2)(\alpha+1)}{}_3F_2\!\left(\begin{matrix}-\gamma,\alpha + \beta+2,1\\ \alpha+\beta+3,\alpha + 2\end{matrix};1\right),
    \end{equation*}
    which can be computed by performing changes of variables and using the identity
    \begin{equation*}
        \begin{split}
            \int_0^1 z^{c-1}(1-z)^{d-1} {_2}F_1(a,b,f; z)\mathrm{d}z &=  B(c,d){_3}F_2(a,b,c;f,c+d;1),
        \end{split}
    \end{equation*}
    where $B(c,d)$ denotes the beta function. 
    
\begin{figure}[!htb]
            \centering
            \includegraphics[width=.5\textwidth]{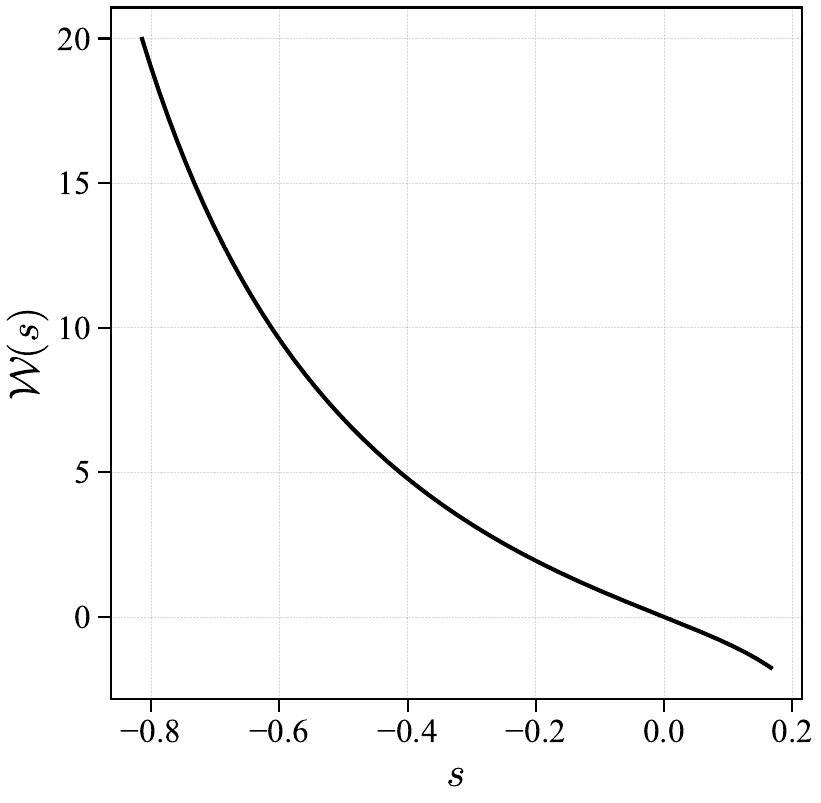}
            \caption{Mellin function along the real axis.}
            \label{fig:Mellin_real}
        \end{figure}

    The Mellin function is analytic in the strip $\Pi(-4/3, 1/6)$ and reads
    \begin{equation*}
        \begin{aligned}
            \Mellin(s)=&-72-\frac12\,{}_3F_2\!\left(\begin{matrix}1,\frac76,\frac43\\ \frac73,\frac52\end{matrix};1\right)-\frac{72}{3s-1}+\frac{216}{6s-1}\\
            &+18\,{}_3F_2\!\left(\begin{matrix}\frac16,1,-s\\ \frac76,\frac43\end{matrix};1\right)
            +18\,{}_3F_2\!\left(\begin{matrix}\frac16,1,\frac76+s\\ \frac76,\frac43\end{matrix};1\right)\\
            &+\frac{1}{1-3s}\Bigg[\frac94\,{}_3F_2\!\left(\begin{matrix}1,\frac76,\frac43\\ \frac73,\frac43-s\end{matrix};1\right)
            -18\,{}_3F_2\!\left(\begin{matrix}1,\frac76,\frac16-s\\ \frac43,\frac76-s\end{matrix};1\right)
            -18\,{}_3F_2\!\left(\begin{matrix}1,\frac16-s,-s\\ \frac43,\frac76-s\end{matrix};1\right)\Bigg]\\
            &+\frac{1}{1-6s}\Bigg[36\,{}_3F_2\!\left(\begin{matrix}1,\frac76,\frac16-s\\ \frac43,\frac76-s\end{matrix};1\right)
            +36\,{}_3F_2\!\left(\begin{matrix}1,\frac16-s,-s\\ \frac43,\frac76-s\end{matrix};1\right)\Bigg]\\
            &+\frac{2}{4-3s}\Bigg[{}_3F_2\!\left(\begin{matrix}1,\frac76,\frac43-s\\ \frac52,\frac73-s\end{matrix};1\right)
            -{}_3F_2\!\left(\begin{matrix}1,\frac76-s,\frac43-s\\ \frac52,\frac73-s\end{matrix};1\right)\Bigg]\\
            &+\frac{1}{4+3s}\Bigg[9\,{}_3F_2\!\left(\begin{matrix}1,\frac76,\frac43+s\\ \frac43,\frac73+s\end{matrix};1\right)
            -2\,{}_3F_2\!\left(\begin{matrix}1,\frac76,\frac43+s\\ \frac52,\frac73+s\end{matrix};1\right)\\
            &\qquad\qquad+9\,{}_3F_2\!\left(\begin{matrix}1,\frac76+s,\frac43+s\\ \frac43,\frac73+s\end{matrix};1\right)
            -2\,{}_3F_2\!\left(\begin{matrix}1,\frac76+s,\frac43+s\\ \frac52,\frac73+s\end{matrix};1\right)\Bigg]\\
            &+\frac{18}{3s+4}\Big[{}_3F_2\!\left(\begin{matrix}1,\frac76,\frac43+s\\ \frac73+s,\frac52+s\end{matrix};1\right)
            +{}_3F_2\!\left(\begin{matrix}1,\frac76+s,\frac43+s\\ \frac73+s,\frac52+s\end{matrix};1\right)\Big]\\
            &-\frac{12}{2s+3}\Big[{}_3F_2\!\left(\begin{matrix}1,\frac76,\frac43+s\\ \frac73+s,\frac52+s\end{matrix};1\right)
            +{}_3F_2\!\left(\begin{matrix}1,\frac76+s,\frac43+s\\ \frac73+s,\frac52+s\end{matrix};1\right)\Big].
        \end{aligned}
    \end{equation*}
    In particular, the Mellin function along the real axis is decreasing, from positive values at $s < 0$ to negative ones for $s > 0$ (Fig.~\ref{fig:Mellin_real}), ensuring that $\Mellin'(0) < 0$.
        Not
\section{Equation for the condensate}
    \label{sec:Condensate}
    Following~\cite{semikoz1995kinetics}, we consider the 4-wave WKE~\eqref{eq:WKE} and decompose the waveaction as 
    \begin{equation*}
        n_{\bm{k}}(t) = \tilde{n}_{\bm{k}}(t) + \delta(\bm{k}) n_c(t),
    \end{equation*}
    where we recall that $\tilde{n}_{\bm{k}}(t)$ corresponds to the "bulk" of the spectrum and $n_c$ to the condensate.
    
    Substituting the above definition into Eq.~\eqref{eq:WKE} yields
    \begin{equation}
        \partial_t \tilde{n}_{\bm{k}} + \delta(\bm{k}) \partial_t n_c = S_4\left[\tilde{n}_{\bm{k}}\right] + 4\pi n_c \int\left[\mathcal{R}^{\bm{k}}_{\bm{2},\bm{3}} - \mathcal{R}^{\bm{2}}_{\bm{k},\bm{3}} - \mathcal{R}^{\bm{3}}_{\bm{2},\bm{k}}\right]\mathrm{d}\bm{k_2}\mathrm{d}\bm{k_3} + 4\pi n_c\delta(\bm{k})\int\mathcal{R}^{\bm{1}}_{\bm{2},\bm{3}}\mathrm{d}\bm{k_1}\mathrm{d}\bm{k_2}\mathrm{d}\bm{k_3},
        \label{eq:cond_deriv}
    \end{equation}
    where $\mathcal{R}^{\bm{k}}_{\bm{2},\bm{3}} = \delta(\bm{k}-\bm{k_2} - \bm{k_3})\delta\omega^k_{2,3}\left[\tilde{n}_{\bm{2}}\tilde{n}_{\bm{3}} - \tilde{n}_{\bm{k}}(\tilde{n}_{\bm{2}}+\tilde{n}_{\bm{3}})\right]$.

    The second term corresponds, up to prefactors, to the 3-wave collision integral obtained for acoustic waves
    \begin{equation*}
        S_3\left[\tilde{n}_{\bm{k}}\right] = 4\pi\int\left[\mathcal{R}^{\bm{k}}_{\bm{2},\bm{3}} - \mathcal{R}^{\bm{2}}_{\bm{k},\bm{3}} - \mathcal{R}^{\bm{3}}_{\bm{2},\bm{k}}\right]\mathrm{d}\bm{k_2}\mathrm{d}\bm{k_3}.
    \end{equation*}
    Integrating $S_3$ over $\bm{k}$ and permuting the indices yields
    \begin{equation*}
        \int S_3\left[\tilde{n}_{\bm{k}}\right] \mathrm{d}\bm{k} = -4\pi \int\mathcal{R}^{\bm{k}}_{\bm{2},\bm{3}}\mathrm{d}\bm{k}\mathrm{d}\bm{k_2}\mathrm{d}\bm{k_3}.
    \end{equation*}
    Substituting into Eq.~\eqref{eq:cond_deriv} leads to
    \begin{equation}
        \partial_t \tilde{n}_{\bm{k}} + \delta(\bm{k}) \partial_t n_c = S_4\left[\tilde{n}_{\bm{k}}\right] + n_cS_3\left[\tilde{n}_{\bm{k}}\right] - n_c\delta(\bm{k})\int S_3\left[\tilde{n}_{\bm{1}}\right]\mathrm{d}\bm{k_1}.
        \label{eq:cond_final}
    \end{equation}
    Interpreting Eq.~\eqref{eq:cond_final} in the sense of distributions in $\mathbf{k}$, we test against $\varphi \in C_c^\infty(\mathbb{R}^3)$ and write
    \begin{equation*}
        \varphi(\mathbf{k}) = \psi(\mathbf{k}) + \varphi(0)\chi(\mathbf{k}),
    \end{equation*}
    where $\psi(0)=0$ and $\chi(0)=1$. 
    
    Substituting into the weak formulation and using the arbitrariness of $\psi$ and $\varphi(0)$, we separate regular and singular contributions to obtain
    \begin{equation*}
        \begin{split}
            \partial_t \tilde{n}_{\mathbf{k}} &= S_4[\tilde{n}_{\mathbf{k}}] + n_c S_3[\tilde{n}_{\mathbf{k}}], \quad \mathbf{k} \neq 0, \\
            \partial_t n_c &= - n_c \int S_3[\tilde{n}_{\mathbf{1}}] \, \mathrm{d}\mathbf{k_1}.
        \end{split}
    \end{equation*}
    Eqs.~\eqref{eq:eq_cond} \& \eqref{eq:bulk} are finally obtained by considering isotropic solutions, averaging over angles~\cite{zhu2026turbulence} and introducing forcing and dissipation in an ad-hoc manner.
\end{document}